\documentclass[superscriptaddress, onecolumn, 11pt]{revtex4-1}

\usepackage{amsmath}    
\usepackage{graphicx}   
\usepackage{verbatim}   
\usepackage{color}      
\usepackage{hyperref}   
\begin{document}

\title{
Data-Driven Sensitivity Inference for Thomson Scattering Electron Density Measurement Systems
}

\author{Keisuke Fujii}
\affiliation{Department of Mechanical Engineering and Science, Graduate School of Engineering, Kyoto University, Kyoto 615-8540, Japan}
\author{Ichihiro Yamada}
\affiliation{National Institute for Fusion Science, Toki 509-5292, Japan}
\author{Masahiro Hasuo}
\affiliation{Department of Mechanical Engineering and Science, Graduate School of Engineering, Kyoto University, Kyoto 615-8540, Japan}

\date{\today}

\begin{abstract}
We developed a method to infer the calibration parameters of multichannel measurement systems, such as channel variations of sensitivity and noise amplitude, from experimental data.
We regard such uncertainties of the calibration parameters as dependent noise.
The statistical properties of the dependent noise and that of the latent functions were modeled and implemented in the Gaussian process kernel. Based on their statistical difference, both parameters were inferred from the data.

We applied this method to the electron density measurement system by Thomson scattering for Large Helical Device plasma, which is equipped with 141 spatial channels.
Based on the 210 sets of experimental data, we evaluated the correction factor of the sensitivity and noise amplitude for each channel.
The correction factor varies by $\approx$ 10\%, and the random noise amplitude is $\approx$ 2\%, i.e., the measurement accuracy increases by a factor of 5 after this sensitivity correction.
The certainty improvement in the spatial derivative inference was demonstrated.

\end{abstract}

\maketitle

\section{Introduction}
Multichannel measurement systems have become more popular in scientific communities because of the reduced cost of digitizers, sensors, and signal processors.
In a multichannel system, sensitivity and noise amplitude variations over channels are important calibration parameters.
Usually, these parameters are evaluated by separate calibration experiments.
However, achieving a highly accurate calibration is sometimes difficult because of the calibration principle or temporal variation of these parameters.
If the sensitivity uncertainty exceeds the signal uncertainty by random statistics,
then the signals will suffer from scatters that are not independent in multiple measurements.

\begin{figure}[h!]
\centering
\includegraphics[scale=0.4]{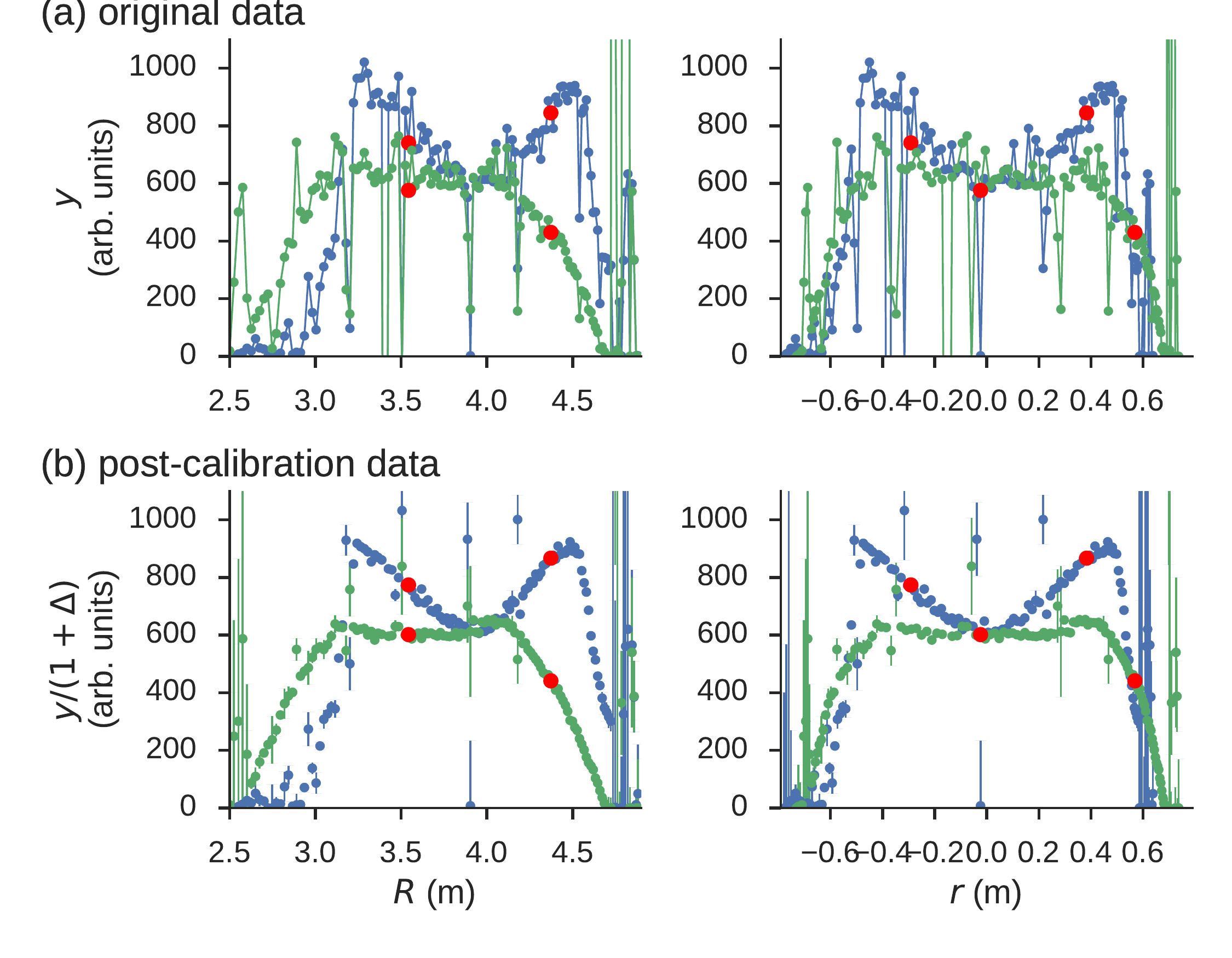}
\caption{
\label{fig:Example-TS}
(a) Two sets of the data provided by the LHD-TS system.
Green points show the data observed for shotnumber \#120725 at $t$ = 4.2 s and blue ones show that for \#120763 at $t$ = 5.6 s.
Some measurement points have a smaller or a larger value than the vicinity points for both results.
The left panel is in the original coordinate (major radius $R$), whereas the right panel is in the mapped coordinate (minor radius $r$).
To clarify the detector channel positions, we show the data points measured by the detector $i$ = 50 and 100 in red markers.
(b) Calibrated results with a miscalibration factor $\Delta$ are inferred in this work. Most scatters were removed after the calibration.
}
\end{figure}

The data observed by a measurement system contains information about the system itself \cite{GPR_application_calibration}.
The upper panel in Fig. \ref{fig:Example-TS} (a) shows two examples of the data observed by one of multichannel systems, specifically,
the electron density ($n_\mathrm{e}$) spatial distribution measurement in the high-temperature plasma (the details of which are provided below).
Some data points have larger or smaller values than those in the vicinity for the two measurements.
This dependent scatter is likely due to the miscalibration of the system sensitivity, and it significantly reduces measurement accuracy.
Humans can clearly distinguish this dependent noise quantitatively by finding the correlation in multiple observation results.
In this study, we demonstrated the machine learning of such dependent noise using the Gaussian process regression (GPR) framework \cite{Rasmussen_GP, Bayesian_analysis}.

GPR is a statistical model that can handle the probability distribution on functions.
The main advantages of GPR are as follows:
\begin{enumerate}
	\item GPR is non-parametric, i.e., a specific function form is not assumed.
	\item GPR is fully based on Bayesian statistics, i.e., if the model complexity is appropriately parameterized, then over- and under-fitting can be avoided.
\end{enumerate}
Given the above advantages, GPR is being adopted in a wide range of fields that involve scientific data analysis,
such as life sciences \cite{GPR_application_life_science1, GPR_application_life_science2},
chemistry \cite{GPR_application_chemistry1, GPR_application_chemistry2},
and
astrophysics \cite{GPR_application_astrophyscis1, GPR_application_astrophyscis2}.
GP has also been used to infer the latent function and its derivatives from data generated by the Thomson scattering (TS) system, which will also be analyzed in this work \cite{TSGP_MAST, TSGP_CMOD}.
However, simple noise has been often assumed for the analysis, e.g., independent homoscedastic Gaussian noise or Poisson noise.
Few works have been reported regarding dependent noise inference \cite{GPR_application_calibration}, which is important for the real-world measurement systems.

In this work, we modeled the statistical properties of dependent noise such as sensitivity variation or channel-dependent noise amplitude into the kernel of Gaussian process (GP).
Using these statistical models, we show that the inference of these noises is straightforward in the GPR framework.
We also applied this technique to the TS system for $n_\mathrm{e}$ spatial distribution measurement in the Large Helical Device (LHD-TS system) \cite{RSI_LHDTS, JINST_LHDTS}, an example of which is shown in Fig. \ref{fig:Example-TS}(a).

The rest of this paper is organized as follows.
In the next section, we briefly describe the fundamental usage of GPR for the latent function and noise inference.
The essence of the inference is based on the statistical difference between the latent function and noise. Therefore, their statistical modeling is important.
We slightly extended the theory in order to treat the dependent noise for the multichannel measurement systems.
In section \ref{sec:application}, we apply this method to the experimental data obtained by the LHD-TS system.
An overview of the system and the statistical model of the noise are described in this section.
The application results are  shown in subsection \ref{subsec:results}, and the effect of this method is presented in section \ref{subsec:applications}, in particular, in the spatial derivative inference.

\section{Dependent Noise Estimation based on GP}

\subsection{Fundamentals of GP \label{subsec:GP_fundamental}}
GP is probability distribution on function space.
If function $f$ follows GP with mean 0 and kernel $\mathrm{K_f}$,
then the joint distribution of any realizations $\mathbf{f} = \{f_i = f(x_i)| i \in 1,...,N\}$ of function $f$ is
\begin{equation}
\label{eq:GP_basics}
p(\mathbf{f}) = \mathcal{N} (0,\mathrm{K_f}),
\end{equation}
where $\mathcal{N}(\mu,\Sigma)$ is the multivariate normal distribution with mean $\mu$ and covariance $\Sigma$ \cite{Rasmussen_GP, Bayesian_analysis, Bishop}.

Additive and independent noise $\mathbf{n}$ is frequently assumed, where the observation data $\mathbf{y} = \{y_i | i\in 1,...,N\}$ are the sum of the
function values and noise, $\mathbf{y} = \mathbf{f} + \mathbf{n}$.
If both $\mathbf{f}$ and $\mathbf{n}$ follow zero mean GP with respective kernels $\mathrm{K_f}$ and $\mathrm{K_n}$,
then the observation and latent function values jointly follow zero mean GP with
\begin{equation}
\label{eq:GP_data}
p\left(
\left[
 \begin{array}{r}
	\mathbf{y} \\
	\mathbf{f}
 \end{array}
\right]\right)
 = \mathcal{N} \left(
 \left[
 \begin{array}{r}
 \mathbf{0} \\
 \mathbf{0}
 \end{array}
 \right]
,
 \left[
 \begin{array}{cc}
 \mathrm{K_f+K_n} & \mathrm{K_f}\\
 \mathrm{K_f} & \mathrm{K_f}
 \end{array}
 \right]\right).
\end{equation}

For the latent function kernel $\mathrm{K_f}$, similarity in the vicinity is often assumed (the meanings of $similar$ and $vicinity$ may depend on problems).
One of the most popular kernels is the radial basis function (RBF) kernel \cite{Rasmussen_GP, Bishop},
\begin{equation}
\label{eq:K_rbf}
\mathrm{K_f}(i,i') = \\
\alpha \exp\left[ - \frac{(x_i-x_{i'})^2}{2l^2} \right]
\end{equation}
where $\mathbf{x} = \{x_i | i \in 1,...,N\}$ is the explanatory variable (usually, spatial coordinates for the multichannel measurement) and $l$ is the scale length.
With this kernel, function values at two close locations, $|x_i - x_{i'}| \ll l$, have a significant correlation of approximately 1,
whereas those at distant locations, $|x_i - x_{i'}| \gg l$, have no correlation.
$\alpha > 0$ exhibits self-variation, which indicates the variation of the function values around zero.

The GP kernel of the additive independent Gaussian noise is written as
\begin{equation}
\label{eq:K_white}
\mathrm{K_n}(i,i') = \\
\sigma^2 \delta(i-i')
\end{equation}
where $\sigma$ is the noise variance, and $\delta(i)$ is the Dirac delta function.

A simple algebraic calculation of Eq. \ref{eq:GP_data} obtains the posterior density of the function values $\mathbf{f}$ and noise $\mathbf{n}$ with given data $\mathbf{y}$ as
\begin{equation}
\label{eq:posterior_f}
p(\mathbf{f}|\mathbf{y}) =
\mathcal{N}\left(
\mathrm{K_f}(\mathrm{K_f}+\mathrm{K_n})^{-1}\mathbf{y}, \mathrm{K_f}-\mathrm{K_f}(\mathrm{K_f}+\mathrm{K_n})^{-1}\mathrm{K_f}
\right),
\end{equation} and
\begin{equation}
\label{eq:posterior_n}
p(\mathbf{n}|\mathbf{y}) =
\mathcal{N}\left(
\mathrm{K_n}(\mathrm{K_f}+\mathrm{K_n})^{-1}\mathbf{y}, \mathrm{K_n}-\mathrm{K_n}(\mathrm{K_f}+\mathrm{K_n})^{-1}\mathrm{K_n}
\right),
\end{equation}
respectively.
This inference is based on the statistical difference of the latent function and noise, which are implemented in the GP kernels, as in Eqs. \ref{eq:K_rbf} and \ref{eq:K_white}.
Note that the notation $p(\mathbf{f}|\mathbf{y})$ represents the conditional probability
distribution of the variable $\mathbf{f}$ conditioned by $\mathbf{y}$.

The set of hyperparameters $\boldsymbol{\theta}$ ($\alpha, l$, and $\sigma$ in this case) can be estimated by the type-II maximum posterior (MAP) method, or they are numerically marginalized out.
In the type-II MAP method, the point estimate of $\boldsymbol{\theta}$ is given by
\begin{equation}
\label{eq:GP_typeIImap}
{\boldsymbol \theta} \approx \mathrm{argmax}_{\boldsymbol \theta} \left[\log p({\boldsymbol \theta}|\mathbf{y})\right]
\end{equation}
where $p({\boldsymbol \theta}|y)$ is the marginal posterior density for ${\boldsymbol \theta}$ with given data $\mathbf{y}$, where $\mathbf{f}$ or $\mathbf{n}$ is marginalized out
\begin{equation}
\label{eq:GP_typeIImap_2}
p({\boldsymbol \theta}|\mathbf{y}) \propto  p(\mathbf{y}|{\boldsymbol \theta}) p({\boldsymbol \theta}),
\end{equation}
where $p({\boldsymbol \theta})$ is the prior distribution of the hyperparameters (hyperprior).
The log of the marginalized likelihood $p(\mathbf{y}|{\boldsymbol \theta})$ can be written as follows \cite{Rasmussen_GP, Bishop}:
\begin{equation}
\label{eq:GP_typeIImap2}
\log p(\mathbf{y}|{\boldsymbol \theta}) =
\log\int  p(\mathbf{y}, \mathbf{f} | {\boldsymbol \theta})\mathrm{d}\mathbf{f} =
-\frac{1}{2}\log|\mathrm{K_f} + \mathrm{K_n} | - \frac{1}{2} \mathbf{y}^\mathsf{T}(\mathrm{K_f} + \mathrm{K_n})^{-1}\mathbf{y} - \frac{N}{2}\log(2\pi).
\end{equation}

\subsection{Common-Mode Noise Inference\label{subsec:GP_commonmode}}
In the previous subsection, we separated the observation data into the latent function component and noise component, as shown in Eqs. \ref{eq:posterior_f} and \ref{eq:posterior_n}, on the basis of their statistical difference.
Such separation is not restricted to the latent function and noise pair.
If data are the sum of multiple components that have different statistical properties and they are appropriately implemented into the GP kernels,
then the posteriors of these elements can be obtained by an equation similar to Eq. \ref{eq:posterior_f} or \ref{eq:posterior_n}.

An example is the common-mode additive noise $\mathbf{n}_\mathrm{C}$, which takes the same values among multiple sets of data (indexed by $j \in 1,...,M$).
Let us consider that data $\mathbf{y}$ are the sum of the latent function values $\mathbf{f}$, the random noise that is independent in each data set $\mathbf{n}_\mathrm{R}$, and the random common-mode noise $\mathbf{n}_\mathrm{C}$.

The kernel of $\mathbf{n}_\mathrm{R}$ for $M$ sets of data can be written as
\begin{equation}
\label{eq:K_white_multi}
\mathrm{K_{n_R}}(i,j,i',j') = \sigma_\mathrm{R}^2 \delta(i-i') \delta(j-j')
\end{equation}
and that for $\mathbf{n}_\mathrm{C}$ is
\begin{equation}
\label{eq:K_common_multi}
\mathrm{K_{n_C}}(i,j,i',j') = \sigma_\mathrm{C}^2 \delta(i-i')
\end{equation}

If the latent functions in the $j$-th data set $\mathbf{f}_j$ are independent from that in another data set, then the kernel of the joint distributions is block diagonal
\begin{equation}
\label{eq:K_f}
\mathrm{K_f} =
\left[
\begin{array}{cccc}
\mathrm{K}_{\mathrm{f}_1} & 0 & \ldots & 0 \\
0 & \mathrm{K}_{\mathrm{f}_2} & \ldots & 0 \\
\vdots &  & \ddots & \vdots \\
0 & 0 & \ldots & \mathrm{K}_{\mathrm{f}_M} \\
\end{array}
\right].
\end{equation}
\noindent
where $\mathrm{K}_{\mathrm{f}_j}$ are the kernels for the latent function in the $j$-th data set.

We assume that all the noise can be approximated to be additive and independent of each other.
Thus, the kernel of the observations is written as $\mathrm{K_y} = \mathrm{K_f} + \mathrm{K_{n_R}} + \mathrm{K_{n_C}} $.
By repeating a discussion similar to Eqs. \ref{eq:posterior_f} and \ref{eq:posterior_n},
we can analytically evaluate the posterior densities of $\mathbf{f}$, $\mathbf{n}_\mathrm{R}$, and $\mathbf{n}_\mathrm{C}$.

Note that the independent assumption of $\mathbf{f}_j$ is too strong because some similarity usually exists among latent functions in multiple data sets.
To protect an artifact from this assumption,
we analyze a few ($M = 6$ in this work) data sets simultaneously.
To analyze more data,
the nondiagonal kernel that represents the similarity needs to be introduced.
Although a bit more details will be discussed in the final section,
it is left for the future study and
we focus on the principle demonstration in this work.

\subsection{Noise Modeling for MultiChannel Systems\label{subsec:GP_dependent}}
In this subsection, we model other realistic dependent noises that are important for multichannel measurement systems.

\subsubsection{Miscalibration Noise}
The sensitivity uncertainty in the multichannel measurement system can be viewed as multiplicative common-mode noise.
Here we define the miscalibration factor for channel $i$ as $\Delta_i$,
where the true calibration factor $R_i$ and the calibration factor that we obtained in the calibration experiment $R_i^0$ are presented in the following relation:
\begin{equation}
\label{eq:calb_factor}
R_i = R_i^0 (1+\Delta_i.).
\end{equation}
We assume $\Delta_i$ is independent of each other and distributed around zero.
One of the prior candidates for $\Delta_i$ is the homoscedastic normal distribution
\begin{equation}
\label{eq:calb_factor_prior}
\Delta_i \sim \mathcal{N}(0, \sigma_{\Delta}^2).
\end{equation}
The noise due to the miscalibration factor is written as $n_{\Delta i,j} = \Delta_i f_{i,j}$.
Therefore, the kernel of this noise $\mathrm{K}_\Delta$ can be approximated as
\begin{equation}
\label{eq:K_delta}
\mathrm{K_{n_\Delta}}(i,i', j,j') = {\sigma_{\Delta}^2} {\hat f}_{i,j}{\hat f}_{i',j'} \delta(i-i'),
\end{equation}
where ${\hat f}_{i,j}$ is an estimate of the latent function value.
We note that Eq. \ref{eq:K_delta} is an approximation because this noise is not strictly additive.
If the posterior width of $f_{i,j}$ is larger than compared to its mean value, then this approximation becomes invalid.

\subsubsection{Channel Dependent Noise}
In real-world multichannel systems, noisy channels that produce more scattered signals than other channels may exist.
We modeled a channel-dependent noise, $\mathbf{n}_\mathrm{D}$, such that the noise variations $\sigma_{\mathrm{D} i}$ depends on channel $i$, while the noise itself varies with zero mean Gaussian distribution.
The kernel of this noise can be written as
\begin{equation}
\label{eq:K_D}
\mathrm{K_{n_D}}(i,i', j,j') = {\sigma_{\mathrm{D} i}^2} \delta(i-i')\delta(j-j'),
\end{equation}
Application of a prior distribution for the noise variances, $p(\sigma_{\mathrm{D}i}^2)$, may be reasonable .

\section{Application to LHD-TS system data \label{sec:application}}





\begin{figure}[h!]
\centering
\includegraphics[scale=0.4]{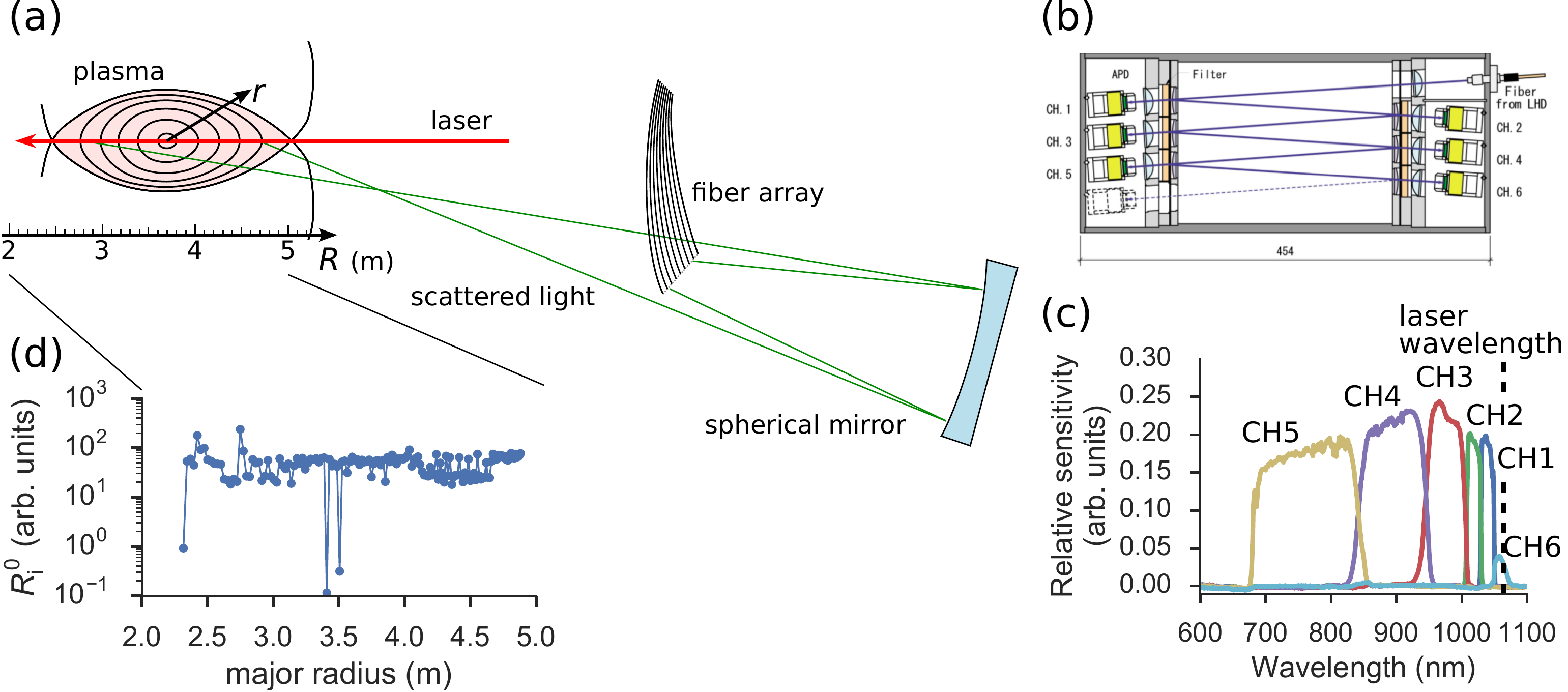}
\caption{
\label{fig:LHD-TS}
Schematic illustration of the LHD-TS system.
(a) Optical configuration of LHD plasma, the incident laser, and an optical system for collecting the scattering light.
The nested magnetic surface, the major radius ($R$), and minor radius ($r$) in the plasma are also shown.
(b) A polychrometer for observing the scattering light spectrum (quoted from Ref. \cite{JINST_LHDTS}).
(c) The relative wavelength sensitivities for the polychrometer measured by the first-step calibration experiment.
(d) The absolute throughput for the LHD-TS system measured by the second-step calibration experiment.
}
\end{figure}

A schematic illustration of the LHD-TS system is shown in Fig. \ref{fig:LHD-TS} \cite{JINST_LHDTS}.
The Nd:YAG laser is injected into the plasma, and the scattered light is focused on the edges of an optical fiber array that comprises 141 fibers [Fig. \ref{fig:LHD-TS} (a)].
The scattering light comes only from the cross-point between the laser path and the sight line,
and therefore this diagnostics system provides high spatial resolution (typically, on the order of a few centimeters).
The temporal resolution is determined by a laser pulse width of ($\approx$ 10 ns), which is much shorter than the time scale of the plasma turbulence.
Given its high spatial and temporal resolution, the TS system has been installed in almost all major magnetic plasma confinement devices \cite{TS_T3, RSI_LHDTS, TS_DIIID, TS_ASDEX, TS_NSTX}.

The light is transferred for $\approx$ 50 m to a set of polychrometers [Fig. \ref{fig:LHD-TS} (b)] in the diagnostics room.
Each polychrometer is equipped with six interference filters and avalanche photodiodes (APD).
The spectral profile of the scattered light is estimated by five channels (CH1--CH5).
The sixth APD (CH6) measures the light around the laser wavelength and
is not used for the measurement because the stray laser light inside the vacuum chamber significantly contaminates this wavelength.

One spectrum is observed by one polychrometer. Therefore, 141 polychrometers are used for the spatial profile measurement.
Although plasma emission is also detected by these detectors,
the intensity of plasma emission is estimated by the subsequent measurement just after the laser pulse and subtracted from the signal \cite{JINST_LHDTS}.

Two-step calibration was conducted for the LHD-TS system.
For the first step, the wavelength-dependent sensitivity of each polychrometer (including the transmittance of the interference filters and sensitivities of the APDs) is calibrated against a standard halogen lamp \cite{JINST_LHDTS}.
An example of the results of the wavelength-dependent sensitivity is shown in Fig. \ref{fig:LHD-TS} (c).
For the second step, the absolute throughput from the scattering position to the APDs is calibrated
by Raman and Rayleigh scattering experiments, which are performed once for each experimental campaign \cite{RSI_LHDTS, PFR_LHDTS}.
The results of the absolute calibration factor $R_i^0$ obtained in the 2013 experimental campaign is shown in Fig. \ref{fig:LHD-TS} (d).
The variation of the calibration factor is almost of an order of magnitude.

The measurement points at each major radius $R$ are mapped on the basis of $T_\mathrm{e}$ results into the nested magnetic flux coordinate, which is denoted by minor radius $r$ [see Fig.\ref{fig:LHD-TS} (a)] \cite{LHD_MAP}.
The mapped results are shown in the right panels in Fig. \ref{fig:Example-TS}.
The parameter profiles, including $n_\mathrm{e}$, are widely accepted to become axisymmetric in the mapped coordinate.

The random independent noise in the LHD-TS data arises because of the finite number of detected photons (shot noise) and thermal noise of the APDs and digitizers,
as well as the fluctuation of the plasma emission.
The dependent noise in the LHD-TS data is due to the miscalibration of the system and the existence of faulty detectors such as those that have bad contact in electric circuits.

\subsection{Statistical Modeling of the LHD-TS System \label{subsec:noise_LHD-TS}}
In this subsection, we describe our modeling of the latent function of $n_\mathrm{e}$ and the noise sources for the LHD-TS system.
For the latent function kernel, $\mathrm{K}_{\mathrm{f}_j}$, we adopted the RBF kernel (Eq. \ref{eq:K_rbf}) and modified it in order to consider the prior information that the $n_\mathrm{e}$ profile in the magnetic plasma confinement device is axisymmetric, as follows:
\begin{equation}
\label{eq:K_rbf_mod}
\mathrm{K_f}(i,i', j,j') = \\
\alpha_j \left(\exp\left[ - \frac{(r_{i,j}-r_{i',j'})^2}{2l_j^2} \right] + \\
\exp\left[ - \frac{(r_{i,j}+r_{i',j'})^2}{2l_j^2} \right] \right)\delta(j-j') .
\end{equation}
where $\alpha_j > 0$ and $l_j > 0$ are hyperparameters for this kernel.
The first term of Eq. \ref{eq:K_rbf_mod} indicates a smooth $n_\mathrm{e}$ profile, i.e., the correlation between two $n_\mathrm{e}$ values that are closer to each other is larger.
The second term makes two $n_\mathrm{e}$ values at the other side against the magnetic axis ($r = 0$) close.
The flat noninformative priors are used for $\alpha_j$ and $l_j$.

We consider the following three types of quasi-additive noise for the LHD-TS data:
\begin{enumerate}
	\item \label{item:mc_noise} Common-mode multiplicative noise, $\mathbf{n}_\Delta$.
	\item \label{item:id_noise} Independent Gaussian noise with a variation that is dependent on channel $\mathbf{n}_\mathrm{D}$.
	\item \label{item:ip_noise} Independent Gaussian noise with a variation proportional to the latent function value, $\mathbf{n}_\mathrm{P}$.
\end{enumerate}
We modeled the kernel of $\mathbf{n}_\Delta$ similar to Eq. \ref{eq:K_delta}, but with a hierarchical form,
\begin{equation}
\label{eq:K_delta2}
\mathrm{K_{n_\Delta}}(i,i', j,j') = {\sigma_{\Delta i}^2} {\hat f}_{i,j}{\hat f}_{i',j'} \delta(i-i'),
\end{equation}
with a hyperprior on $\sigma_{\Delta i}^2$,
\begin{equation}
\label{eq:hyperprior_mc}
\sigma_{\mathrm{\Delta}i}^2 \sim \mathcal{IG}(\frac{1}{2},\frac{\bar \sigma_\mathrm{\Delta}^2}{2}).
\end{equation}
where $\mathcal{IG}(\alpha, \beta)$ is the inverse gamma distribution with a shape parameter $\alpha$ and scale parameter $\beta$.
Moreover, the variance of the hyperprior $\bar \sigma_\mathrm{\Delta}^2$ is another hyperparameter.
Once the variance $\sigma_{\mathrm{\Delta} i}$ is marginalized out, the distribution of this noise becomes Cauchy distribution with scale $\bar \sigma_\mathrm{\Delta}$.

For the GP kernel of $\mathbf{n}_\mathrm{D}$, we adopted Eq. \ref{eq:K_D} also with the prior distribution on $\sigma_{\mathrm{D}i}$
\begin{equation}
\label{eq:hyperprior_id}
\sigma_{\mathrm{D}i}^2 \sim \mathcal{IG}(\frac{1}{2},\frac{\bar \sigma_\mathrm{D}^2}{2}).
\end{equation}

We modeled the third noise $\mathbf{n}_\mathrm{P}$ such that its variation is proportional to the latent function value.
The GP kernel for this noise can be approximated by
\begin{equation}
\label{eq:kernel_ip}
\mathrm{K_{n_P}}(i,i', j,j') = {\sigma_\mathrm{P}^2} {\hat f}_{i,j}^2 \delta(i-i')\delta(j-j').
\end{equation}

The above modeling is considerably simplified, especially, for low $T_\mathrm{e}$ plasmas.
The spectral profile of the scattered light from a low $T_\mathrm{e}$ plasma, which is usually located at the edges of plasma, is close to the laser spectrum.
The CH6 APD in Fig. \ref{fig:LHD-TS}(c) cannot be used for the measurement due to the strong stray light.
Thus, the evaluation of $T_\mathrm{e}$ and $n_\mathrm{e}$ values from the small $T_\mathrm{e}$ plasma becomes unstable, and the evaluated values sometimes deviate significantly from the true value.

In our modeling, this $T_\mathrm{e}$ dependence of the signal variance is not considered.
We model only the channel dependence of the variation. Thus, the noise variation of the channels observed on the edges of plasma is inferred to be very large.
However, we focus on the evaluation of the miscalibration factor in this work.

The point estimates of the hyperparameters in these kernels were conducted by the type-II MAP method (Eq. \ref{eq:GP_typeIImap_2}).
Because kernels contain ${\hat f}_{i,j}$, we performed an iterative evaluation until the solutions converged.

\subsection{Results \label{subsec:results}}
We prepared 210 sets of LHD-TS data (from two LHD discharges, \#120724 and \#120762)
in the 2013 experimental campaign.
The number of spatial channels $N$ is 141.
The radial coordinate $r$ is mapped based on the basis of VMEC calculation \cite{LHD_MAP}.

We rondomly divided 210 data set into 35 mini-batches where
in each mini-batch we analyzed $M=6$ data sets at once.
The results were combined with a Bayesian committee machine \cite{BCM},
\begin{equation}
\label{eq:BCM}
p(\mathbf{n}_\Delta|\mathbf{y}^{(1)},\cdots,\mathbf{y}^{(D)}) \propto
\prod_{d=1}^{D} \frac{p(\mathbf{n}_\Delta|\mathbf{y}^{(d)})}{p(\mathbf{n}_\Delta)^{D-1}},
\end{equation}
where $\mathbf{y}^{(d)}$ is the $d$-th mini-batch indexed
by $d = {1,\cdots,D}$.
In this work, $D = 35$.
Note that

Such mini-batch calculation is adopted not only to relax the large calculation cost of GP, i.e., $\mathcal{O}((N\times M)^3)$,
but also to avoid the strong assumption in Eq. \ref{eq:K_f}.

\begin{figure}[h!]
\centering
\includegraphics[scale=0.4]{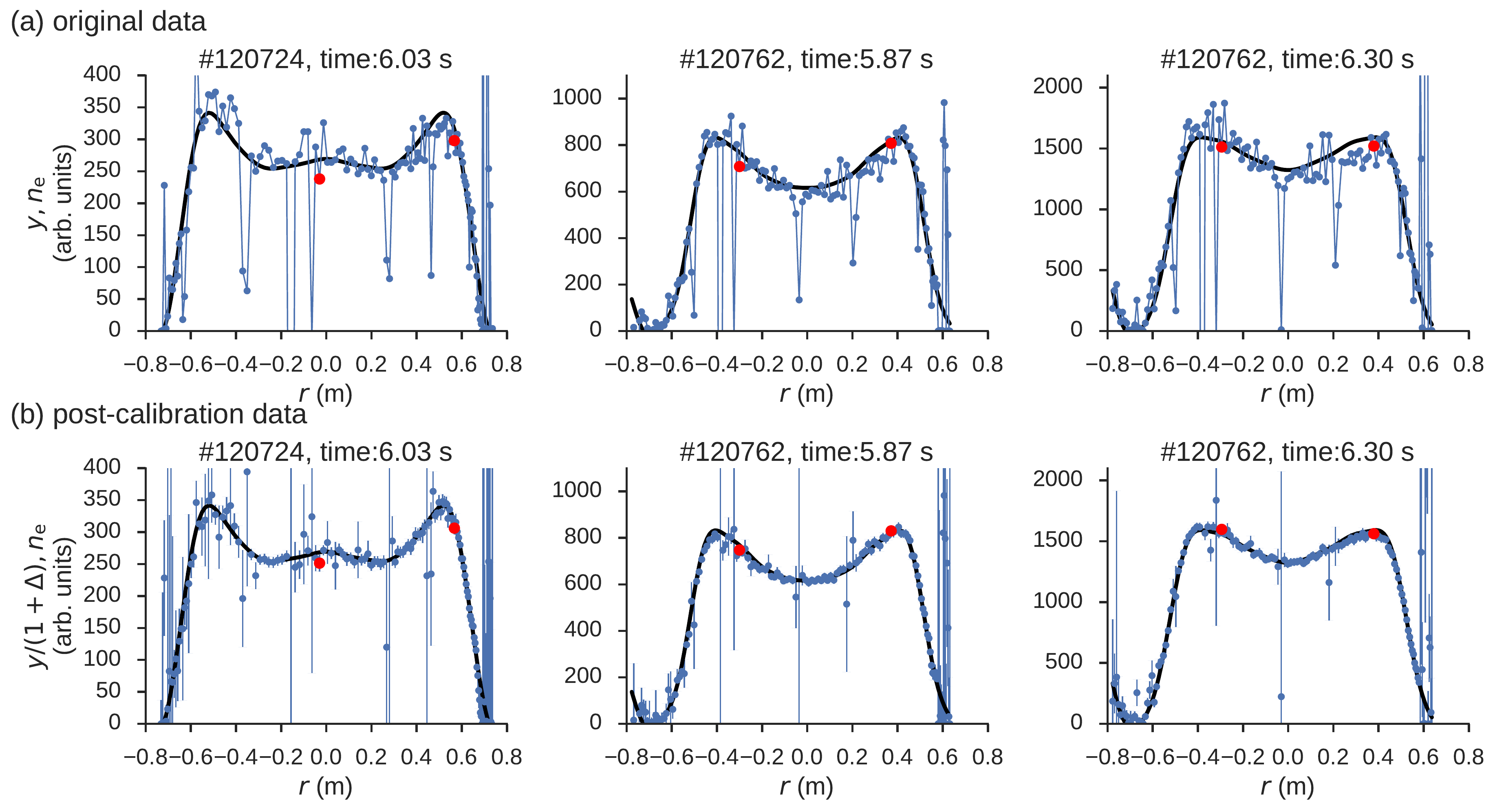}
\caption{
\label{fig:ne_result}
(a) Examples of the data $\mathbf{y}$ obtained by the LHD-TS system as a function of effective minor radius $r$.
To show the mapping effect for each data set, the data obtained by channels $i = $ 50 and 100 are shown in red.
(b) Calibrated values, $y_i/(1+\Delta_i)$. Scatters of the signals are significantly removed.
(Black curves) Posterior mean of the latent functions $f_{i,j}$.
}
\end{figure}

\begin{figure}
	\centering
	\includegraphics[scale=0.6]{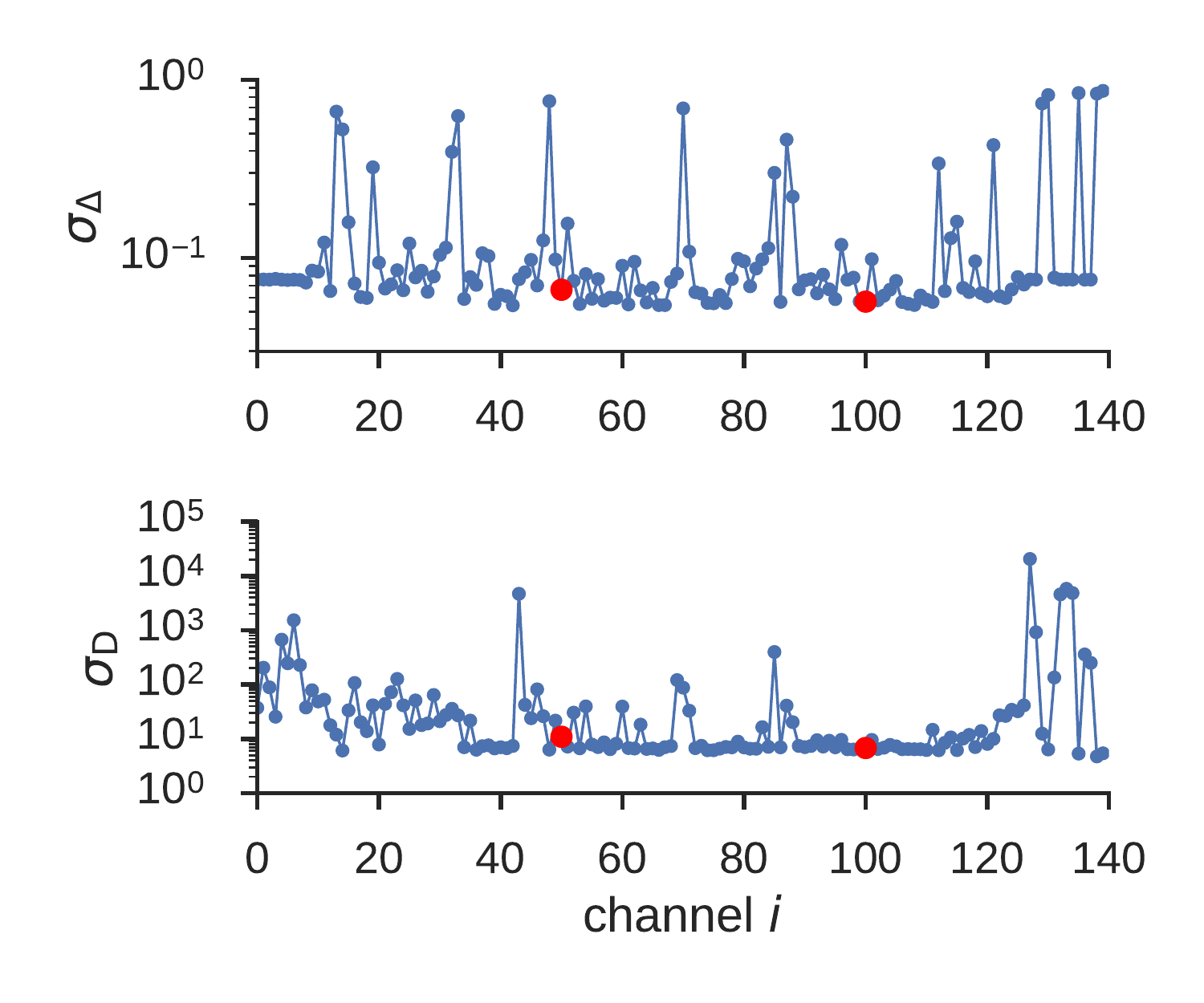}
	\caption{
		\label{fig:sigma_result}
		The variance of miscalibration noise $\sigma_{\Delta i}$ and
		the random noise variance $\sigma_{\mathrm{D}i}$ that maximize the marginal posterior Eq. \ref{eq:GP_typeIImap}.
		Results for channels $i = $ 50 and 100 are shown in red.
	}
\end{figure}

\begin{figure}[h!]
\centering
\includegraphics[scale=0.6]{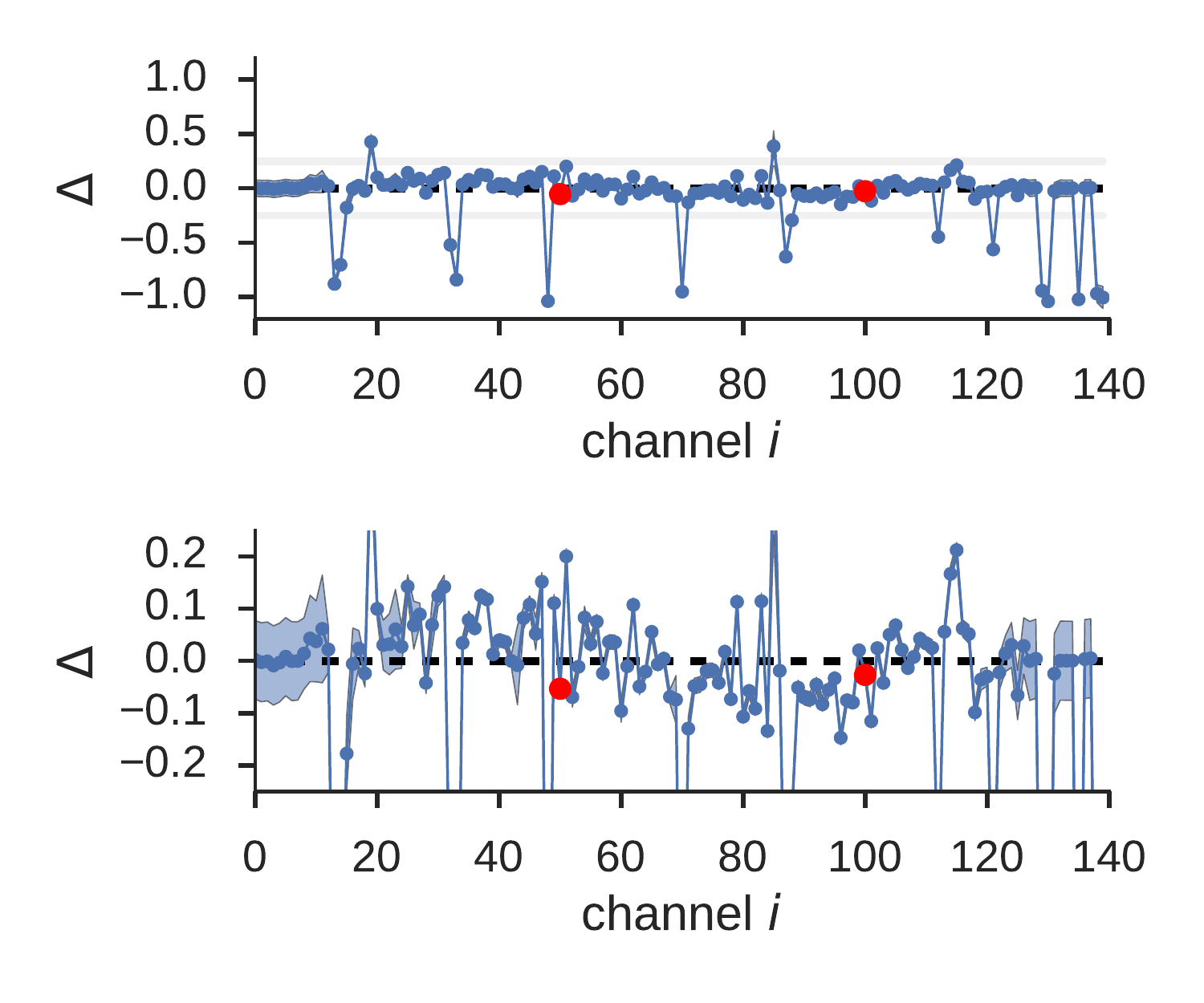}
\caption{
\label{fig:delta_result}
Miscalibration noise $\Delta_i$ inferred from one mini-batch.
The posterior mean is shown by markers and the range within one standard deviation are shown with shadow.
The values for channels $i = $ 50 and 100 are shown in red.
The lower panel is the vertical expansion of the upper panel.
}
\end{figure}

Figure \ref{fig:ne_result} (a) shows some of the experimental data used for one mini-batch.
We maximized the marginalized posterior conditioned by this data set (Eq. \ref{eq:GP_typeIImap2}).
Moreover, $\sigma_{\Delta i}$ and $\sigma_{\mathrm{D}i}$ that maximize the posterior are shown in Fig. \ref{fig:sigma_result}.
Higher $\sigma_{\mathrm{D} i}$ values were evaluated for possible faulty channels, e.g. $i = 43$, and both ends of the plasma $i < 30$ and $i > 123$ where the $T_\mathrm{e}$ values are always low.
${\bar{\sigma}_{\Delta}}$, ${\bar{\sigma}_{\mathrm{D}}}$, and ${\sigma_\mathrm{P}}$ were evaluated as 0.075, 9.3, and 0.018, respectively.

With these hyperparameters, we evaluated the posterior distributions for the latent functions for $n_\mathrm{e}$ and the miscalibration factor $\Delta_i$ based on Eqs. \ref{eq:posterior_f} and \ref{eq:posterior_n}, respectively.
The posteriors of $n_\mathrm{e}$ and $\Delta_i$ are shown in Fig. \ref{fig:ne_result} by black curves and in Fig. \ref{fig:delta_result} by blue points with shadows that represent their one standard deviations, respectively.

Based on a relation $y_{i,j} = f_{i,j} + n_{\mathrm{D} i,j} + n_{\mathrm{P} i,j} + \Delta_i {\hat f}_{i,j}$,
the following calibration can be made with the given $\Delta_i $ values:
\begin{equation}
\label{eq:f_approx}
\frac{y_{i,j}}{1+\Delta_i} = f_{i,j} + \frac{n_{\mathrm{D} i,j} + n_{\mathrm{P}i,j} }{1+\Delta_i}.
\end{equation}
The mean and standard deviation of $y_{i,j}/(1+\Delta_i)$ are shown in Fig. \ref{fig:ne_result} (b) by blue markers and error bars.
The scatters of the data points are largely suppressed by this post-calibration.

\begin{figure}
\centering
\includegraphics[scale=0.6]{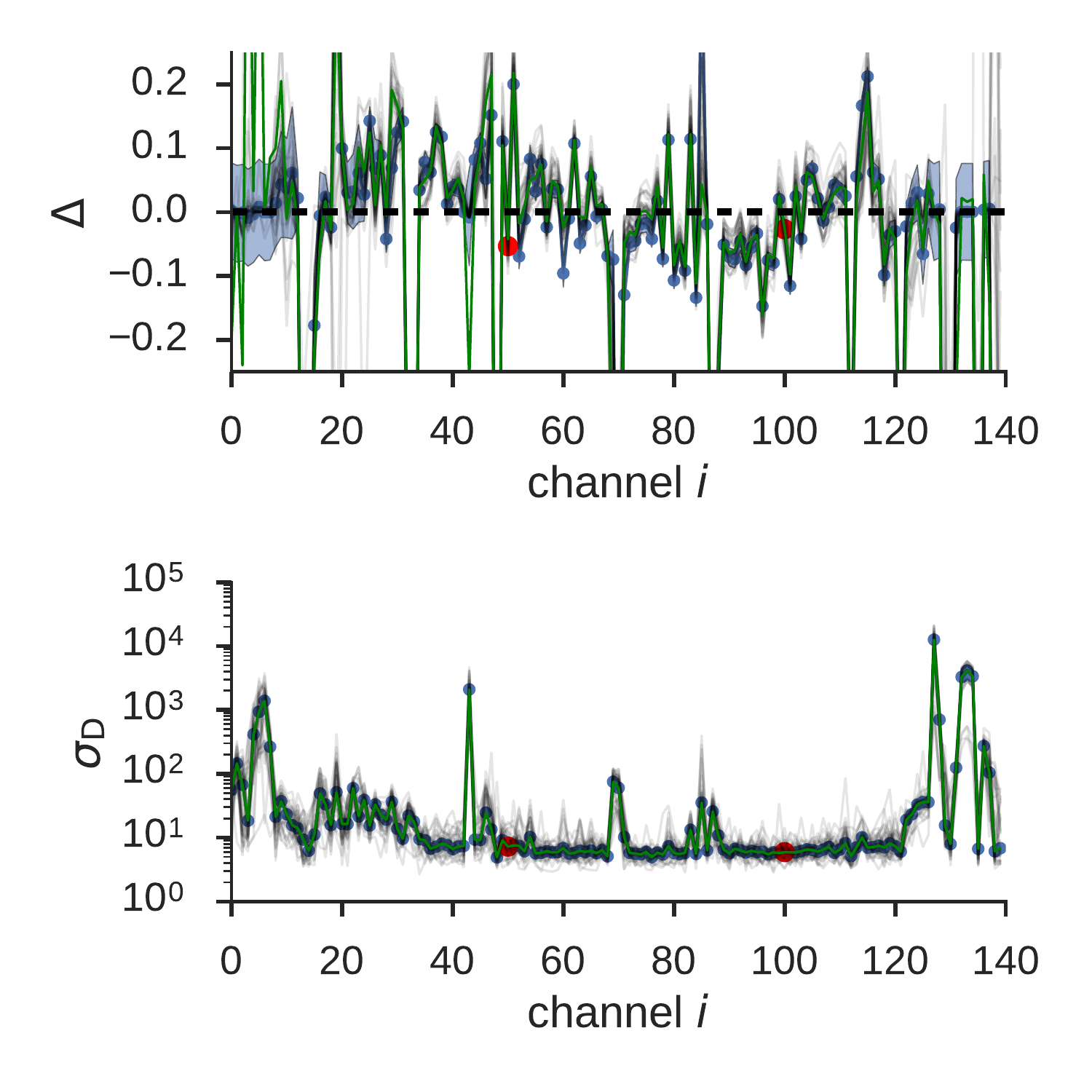}
\caption{
\label{fig:sigma_delta_batch}
(Upper panel)
Light gray curves show the posterior means of Miscalibration noise $\Delta$
for all the data sets.
The green curves show the combined result by the Bayesian committee machine.
Blue markers and blue shadow are the evaluated results for one small data set,
which is the same as those in Figs.\ref{fig:delta_result}
(Lower panel)
Light gray curves show point estimates of random noise variance $\sigma_\mathrm{D}$,
for all the data sets.
The green curve is the median of them.
Blue markers are the evaluated results for one small data set,
which is the same as those in Figs. \ref{fig:sigma_result}.
}
\end{figure}

The upper and lower panels of Fig. \ref{fig:sigma_delta_batch} show the posterior mean of $\Delta_i$ and $\sigma_{\mathrm{D}i}$ for the 35 mini-batches, respectively.
These results have similar values.
The green line in the upper panel of Fig. \ref{fig:sigma_delta_batch} shows the combined results of $\Delta_i$ by the Bayesian committee machine,
while that in the lower panel is the median of these results of $\sigma_{\mathrm{D} i}$.
Note that the Bayesian committee machine is not used to combine multiple results
of $\sigma_{\mathrm{D} i}$ since we only make a point estimate of this hyperparameter
rather than its posterior distribution.

With these combined results, we calibrated test data that are not used for the above inference.
The calibration was performed on the basis of Eq. \ref{eq:f_approx}.
Some of the original data and the results calibrated with $\Delta_i$ values (post-calibration) are shown in Figs. \ref{fig:Example-TS} (a) and (b), respectively.
Most scatters in the test data were also removed, thereby indicating
 that our method successfully avoids overfitting.
A few outliers still exist in the post-calibrated data, and
they are likely due to the simplified modeling for the outliers.

\subsection{Spatial Derivative Inference \label{subsec:applications}}
We demonstrate a benefit of this post-calibration by evaluating the latent function and its first and second spatial derivatives,
which are important for the plasma simulations and transport analysis \cite{SIMULATION1}.

\begin{figure}[h!]
\centering
\includegraphics[scale=0.5]{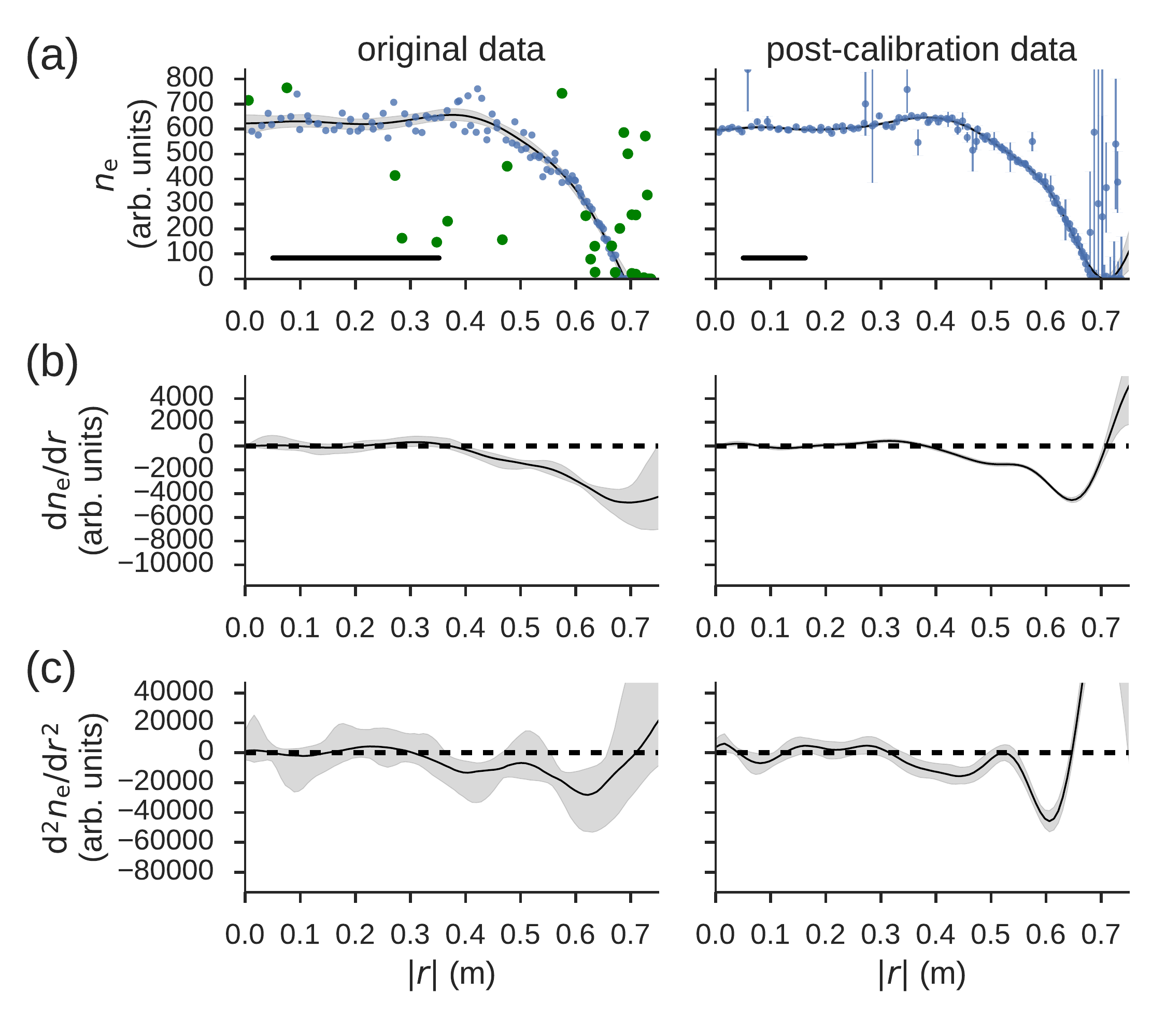}
\caption{
\label{fig:application1}
Analyzed results for the test data (shot number = 120725, $t = $4.2 s).
(a) The left panel shows the original data measured by the LHD-TS system. The green markers are outliers that are chosen manually and removed for the inference.
The right panel shows the post-calibrated data.
The posterior median of the inferred latent functions by MCMC are shown by black curves.
The horizontal bar in each panel shows the length scale in the RBF kernel that maximizes the marginal posterior.
(b) First and (c) second spatial derivatives of the latent functions.
The range within 95\% intervals of the posterior is shown with shadows in each panel.
}
\end{figure}

A original test data set analyzed here is shown in the left panel of Fig. \ref{fig:application1} (a)
which is also shown by the green markers in Fig. \ref{fig:Example-TS} (a).
The post-calibration was performed on the basis of Eq. \ref{eq:f_approx} for this test data set, and the result is shown in the right panel.

To analyze the original data (without the post-calibration), we adopted a basic GP regression
\begin{equation}
\label{eq:K_test_original}
\mathrm{K} = \mathrm{K_f} + \mathrm{K}_\mathrm{n_{R}},
\end{equation}
as described in subsection \ref{subsec:GP_fundamental}.
The basic GP regression cannot deal with the outliers. We manually removed the outliers from the analysis as shown by green markers in Fig. \ref{fig:application1} from the analysis.

For the post-calibration data, we adopted the following kernel,
\begin{equation}
\label{eq:K_test_calb}
\mathrm{K} = \mathrm{K_f} + \mathrm{K_{n_D}} + \mathrm{K_{n_P}},
\end{equation}
where the hyperparameters in $\mathrm{K_{n_C}}$ and $\mathrm{K_{n_P}}$ are fixed to those evaluated for the training data multiplied by $1/(1+\Delta_i)$.
We have already inferred outliers as signal variation $\sigma_{\mathrm{D} i}$. Manual removal of the outliers is not necessary
for the post-calibration data.

Here we carried out Markov chain Monte-Carlo (MCMC) method to marginalize
all the hyperparameters.
The inferred latent functions are shown by black curves in Fig. \ref{fig:application1} (a) for both cases.
The scale length of the RBF kernel ($l$ in Eq. \ref{eq:K_rbf_mod}) that maximizes the mearginal posterior is shown by the horizontal bar.
The ranges within two standard deviation of the posterior are shown in gray shadow.
The optimum scale length for the original test data in Fig. \ref{fig:application1} is larger than that for the post-calibrated data,
resulting in lower spatial resolution for the inference.

We also evaluated the spatial derivatives against the minor radius $r$ by GP \cite{Rasmussen_GP,TSGP_CMOD}.
The inferred first and second derivatives are shown in (b) and (c) in Fig.
\ref{fig:application1}, respectively.
The spatial resolution (inverse of the scale length) or the certainties of the derivative evaluation
also increase by the post-calibration.
The spatial resolution (inverse of the scale length) or the certainties of the derivative evaluation also increase by the post-calibration.

\section{Summary and Future prospects}
We developed a method to infer the miscalibration factor and channel dependent noise amplitude for multichannel measurement systems based on GPR.
In this method, the dependent noises are statistically modeled in the kernel that is hierarchically parameterized.
These hyperparameters were estimated so that they jointly maximize the marginal posterior.
The posterior distributions of the miscalibration factor were inferred with these hyperparameters.
We applied this method to the TS system for $n_\mathrm{e}$ measurement in LHD.
We showed that the post-calibration with these miscalibration factors and channel-dependent noise amplitude
significantly decreases the uncertainty, especially, for spatial derivative estimation.

In the present stage of analysis, we adopted the mini-batch calculation with
$M = 6$, to avoid the artifact due to the too simple assumption, Eq.\ref{eq:K_f}.
Equation \ref{eq:K_f} is not valid for analyizing many sets of data,
since this assumes that $M$ sets of the data are random samples from
$\mathcal{N}(0,\mathrm{K_f})$, while there are some clear similarities
(or correlations) among them,
e.g. all the $n_\mathrm{e}$ profiles are positive-valued and
have larger values in the plasma axis than that at the edge region.
In the future analysis, it is necessary to adopt more appropriate models that
distinguishes the correlation among latent functions and the dependent noise.
A possible model that takes the correlation of the latent function
 into account is the coregionalized model \cite{MultitaskGP}, which models non-diagonal covariance matrix.

Another future improvement is to consider the $T_\mathrm{e}$ dependence of
the noise variance, which we neglected in this work.
Due to the spectroscopic configuration of the LHD-TS system,
the measurement accurary for $n_\mathrm{e}$ decreases in very low or high $T_\mathrm{e}$ plasmas.
The measurement accuracy for $T_\mathrm{e}$ also depends on both
$T_\mathrm{e}$ and $n_\mathrm{e}$ values.
Such correlation should be taken into account to infer the latent functions
in wide range of plasma parameters.

\begin{acknowledgements}
The authors thank the development team of pyGPs, the Python implementation of GP.
The authors would also like to thank the LHD Experiment Group for the smooth LHD operation. Part of this work was supported by JSPS KAKENHI grant number 26610191, the National Institute for Fusion Science (NIFS14KLPH023, NIFS13KLPF032).

\end{acknowledgements}

\bibliography{refs}

\begin{thebibliography}{23}%
\makeatletter
\providecommand \@ifxundefined [1]{%
 \@ifx{#1\undefined}
}%
\providecommand \@ifnum [1]{%
 \ifnum #1\expandafter \@firstoftwo
 \else \expandafter \@secondoftwo
 \fi
}%
\providecommand \@ifx [1]{%
 \ifx #1\expandafter \@firstoftwo
 \else \expandafter \@secondoftwo
 \fi
}%
\providecommand \natexlab [1]{#1}%
\providecommand \enquote  [1]{``#1''}%
\providecommand \bibnamefont  [1]{#1}%
\providecommand \bibfnamefont [1]{#1}%
\providecommand \citenamefont [1]{#1}%
\providecommand \href@noop [0]{\@secondoftwo}%
\providecommand \href [0]{\begingroup \@sanitize@url \@href}%
\providecommand \@href[1]{\@@startlink{#1}\@@href}%
\providecommand \@@href[1]{\endgroup#1\@@endlink}%
\providecommand \@sanitize@url [0]{\catcode `\\12\catcode `\$12\catcode
  `\&12\catcode `\#12\catcode `\^12\catcode `\_12\catcode `\%12\relax}%
\providecommand \@@startlink[1]{}%
\providecommand \@@endlink[0]{}%
\providecommand \url  [0]{\begingroup\@sanitize@url \@url }%
\providecommand \@url [1]{\endgroup\@href {#1}{\urlprefix }}%
\providecommand \urlprefix  [0]{URL }%
\providecommand \Eprint [0]{\href }%
\providecommand \doibase [0]{http://dx.doi.org/}%
\providecommand \selectlanguage [0]{\@gobble}%
\providecommand \bibinfo  [0]{\@secondoftwo}%
\providecommand \bibfield  [0]{\@secondoftwo}%
\providecommand \translation [1]{[#1]}%
\providecommand \BibitemOpen [0]{}%
\providecommand \bibitemStop [0]{}%
\providecommand \bibitemNoStop [0]{.\EOS\space}%
\providecommand \EOS [0]{\spacefactor3000\relax}%
\providecommand \BibitemShut  [1]{\csname bibitem#1\endcsname}%
\let\auto@bib@innerbib\@empty
\bibitem [{\citenamefont {Gibson}\ \emph {et~al.}(2012)\citenamefont {Gibson},
  \citenamefont {Aigrain}, \citenamefont {Roberts}, \citenamefont {Evans},
  \citenamefont {Osborne},\ and\ \citenamefont
  {Pont}}]{GPR_application_calibration}%
  \BibitemOpen
  \bibfield  {author} {\bibinfo {author} {\bibfnamefont {N.~P.}\ \bibnamefont
  {Gibson}}, \bibinfo {author} {\bibfnamefont {S.}~\bibnamefont {Aigrain}},
  \bibinfo {author} {\bibfnamefont {S.}~\bibnamefont {Roberts}}, \bibinfo
  {author} {\bibfnamefont {T.~M.}\ \bibnamefont {Evans}}, \bibinfo {author}
  {\bibfnamefont {M.}~\bibnamefont {Osborne}}, \ and\ \bibinfo {author}
  {\bibfnamefont {F.}~\bibnamefont {Pont}},\ }\href {\doibase
  10.1111/j.1365-2966.2011.19915.x} {\bibfield  {journal} {\bibinfo  {journal}
  {Monthly Notices of the Royal Astronomical Society}\ }\textbf {\bibinfo
  {volume} {419}},\ \bibinfo {pages} {2683} (\bibinfo {year} {2012})},\ \Eprint
  {http://arxiv.org/abs/http://mnras.oxfordjournals.org/content/419/3/2683.full.pdf+html}
  {http://mnras.oxfordjournals.org/content/419/3/2683.full.pdf+html}
  \BibitemShut {NoStop}%
\bibitem [{\citenamefont {Rasmussen}\ and\ \citenamefont
  {Williams}(2005)}]{Rasmussen_GP}%
  \BibitemOpen
  \bibfield  {author} {\bibinfo {author} {\bibfnamefont {C.~E.}\ \bibnamefont
  {Rasmussen}}\ and\ \bibinfo {author} {\bibfnamefont {C.~K.~I.}\ \bibnamefont
  {Williams}},\ }\href@noop {} {\emph {\bibinfo {title} {Gaussian Processes for
  Machine Learning}}}\ (\bibinfo  {publisher} {MIT Press},\ \bibinfo {address}
  {US One Rogers Street Cambridge MA 02142-1209},\ \bibinfo {year}
  {2005})\BibitemShut {NoStop}%
\bibitem [{\citenamefont {Gelman~A.}(2013)}]{Bayesian_analysis}%
  \BibitemOpen
  \bibfield  {author} {\bibinfo {author} {\bibfnamefont {S.~H. S. D. D. B. V.
  A. R. D.~B.}\ \bibnamefont {Gelman~A.}, \bibfnamefont {Carlin J.~B.}},\
  }\href@noop {} {\emph {\bibinfo {title} {Bayesian Data Analysis 3rd edn}}}\
  (\bibinfo  {publisher} {CRC Press},\ \bibinfo {address} {6000 Broken Sound
  Pkwy NW 300, Boca Raton, FL 33487},\ \bibinfo {year} {2013})\BibitemShut
  {NoStop}%
\bibitem [{\citenamefont {Stegle$^{*}$}\ \emph {et~al.}(2008)\citenamefont
  {Stegle$^{*}$}, \citenamefont {Fallert}, \citenamefont {MacKay},\ and\
  \citenamefont {Brage}}]{GPR_application_life_science1}%
  \BibitemOpen
  \bibfield  {author} {\bibinfo {author} {\bibfnamefont {O.}~\bibnamefont
  {Stegle$^{*}$}}, \bibinfo {author} {\bibfnamefont {S.~V.}\ \bibnamefont
  {Fallert}}, \bibinfo {author} {\bibfnamefont {D.~J.~C.}\ \bibnamefont
  {MacKay}}, \ and\ \bibinfo {author} {\bibfnamefont {S.}~\bibnamefont
  {Brage}},\ }\href {\doibase 10.1109/TBME.2008.923118} {\bibfield  {journal}
  {\bibinfo  {journal} {IEEE Transactions on Biomedical Engineering}\ }\textbf
  {\bibinfo {volume} {55}},\ \bibinfo {pages} {2143} (\bibinfo {year}
  {2008})}\BibitemShut {NoStop}%
\bibitem [{\citenamefont {Kemmler}\ \emph {et~al.}(2013)\citenamefont
  {Kemmler}, \citenamefont {Rodner}, \citenamefont {R^^c3^^b6sch},
  \citenamefont {Popp},\ and\ \citenamefont
  {Denzler}}]{GPR_application_life_science2}%
  \BibitemOpen
  \bibfield  {author} {\bibinfo {author} {\bibfnamefont {M.}~\bibnamefont
  {Kemmler}}, \bibinfo {author} {\bibfnamefont {E.}~\bibnamefont {Rodner}},
  \bibinfo {author} {\bibfnamefont {P.}~\bibnamefont {R^^c3^^b6sch}}, \bibinfo
  {author} {\bibfnamefont {J.}~\bibnamefont {Popp}}, \ and\ \bibinfo {author}
  {\bibfnamefont {J.}~\bibnamefont {Denzler}},\ }\href {\doibase
  http://dx.doi.org/10.1016/j.aca.2013.07.051} {\bibfield  {journal} {\bibinfo
  {journal} {Analytica Chimica Acta}\ }\textbf {\bibinfo {volume} {794}},\
  \bibinfo {pages} {29 } (\bibinfo {year} {2013})}\BibitemShut {NoStop}%
\bibitem [{\citenamefont {Burden†}(2001)}]{GPR_application_chemistry1}%
  \BibitemOpen
  \bibfield  {author} {\bibinfo {author} {\bibfnamefont {F.~R.}\ \bibnamefont
  {Burden†}},\ }\href {\doibase 10.1021/ci000459c} {\bibfield  {journal}
  {\bibinfo  {journal} {Journal of Chemical Information and Computer Sciences}\
  }\textbf {\bibinfo {volume} {41}},\ \bibinfo {pages} {830} (\bibinfo {year}
  {2001})},\ \bibinfo {note} {pMID: 11410065},\ \Eprint
  {http://arxiv.org/abs/http://dx.doi.org/10.1021/ci000459c}
  {http://dx.doi.org/10.1021/ci000459c} \BibitemShut {NoStop}%
\bibitem [{\citenamefont {Chen}\ \emph {et~al.}(2007)\citenamefont {Chen},
  \citenamefont {Morris},\ and\ \citenamefont
  {Martin}}]{GPR_application_chemistry2}%
  \BibitemOpen
  \bibfield  {author} {\bibinfo {author} {\bibfnamefont {T.}~\bibnamefont
  {Chen}}, \bibinfo {author} {\bibfnamefont {J.}~\bibnamefont {Morris}}, \ and\
  \bibinfo {author} {\bibfnamefont {E.}~\bibnamefont {Martin}},\ }\href
  {\doibase http://dx.doi.org/10.1016/j.chemolab.2006.09.004} {\bibfield
  {journal} {\bibinfo  {journal} {Chemometrics and Intelligent Laboratory
  Systems}\ }\textbf {\bibinfo {volume} {87}},\ \bibinfo {pages} {59 }
  (\bibinfo {year} {2007})},\ \bibinfo {note} {selected papers presented at the
  Conferentia Chemometrica 2005 Hajd^^c3^^baszoboszl^^c3^^b3, Hungary 28-31
  August 2005Conferentia Chemometrica 2005}\BibitemShut {NoStop}%
\bibitem [{\citenamefont {Way}\ \emph {et~al.}(2009)\citenamefont {Way},
  \citenamefont {Foster}, \citenamefont {Gazis},\ and\ \citenamefont
  {Srivastava}}]{GPR_application_astrophyscis1}%
  \BibitemOpen
  \bibfield  {author} {\bibinfo {author} {\bibfnamefont {M.~J.}\ \bibnamefont
  {Way}}, \bibinfo {author} {\bibfnamefont {L.~V.}\ \bibnamefont {Foster}},
  \bibinfo {author} {\bibfnamefont {P.~R.}\ \bibnamefont {Gazis}}, \ and\
  \bibinfo {author} {\bibfnamefont {A.~N.}\ \bibnamefont {Srivastava}},\ }\href
  {http://stacks.iop.org/0004-637X/706/i=1/a=623} {\bibfield  {journal}
  {\bibinfo  {journal} {The Astrophysical Journal}\ }\textbf {\bibinfo {volume}
  {706}},\ \bibinfo {pages} {623} (\bibinfo {year} {2009})}\BibitemShut
  {NoStop}%
\bibitem [{\citenamefont {Bonfield}\ \emph {et~al.}(2010)\citenamefont
  {Bonfield}, \citenamefont {Sun}, \citenamefont {Davey}, \citenamefont
  {Jarvis}, \citenamefont {Abdalla}, \citenamefont {Banerji},\ and\
  \citenamefont {Adams}}]{GPR_application_astrophyscis2}%
  \BibitemOpen
  \bibfield  {author} {\bibinfo {author} {\bibfnamefont {D.~G.}\ \bibnamefont
  {Bonfield}}, \bibinfo {author} {\bibfnamefont {Y.}~\bibnamefont {Sun}},
  \bibinfo {author} {\bibfnamefont {N.}~\bibnamefont {Davey}}, \bibinfo
  {author} {\bibfnamefont {M.~J.}\ \bibnamefont {Jarvis}}, \bibinfo {author}
  {\bibfnamefont {F.~B.}\ \bibnamefont {Abdalla}}, \bibinfo {author}
  {\bibfnamefont {M.}~\bibnamefont {Banerji}}, \ and\ \bibinfo {author}
  {\bibfnamefont {R.~G.}\ \bibnamefont {Adams}},\ }\href {\doibase
  10.1111/j.1365-2966.2010.16544.x} {\bibfield  {journal} {\bibinfo  {journal}
  {Monthly Notices of the Royal Astronomical Society}\ }\textbf {\bibinfo
  {volume} {405}},\ \bibinfo {pages} {987} (\bibinfo {year} {2010})},\ \Eprint
  {http://arxiv.org/abs/http://mnras.oxfordjournals.org/content/405/2/987.full.pdf+html}
  {http://mnras.oxfordjournals.org/content/405/2/987.full.pdf+html}
  \BibitemShut {NoStop}%
\bibitem [{\citenamefont {von Nessi}\ and\ \citenamefont
  {Hole}(2013)}]{TSGP_MAST}%
  \BibitemOpen
  \bibfield  {author} {\bibinfo {author} {\bibfnamefont {G.~T.}\ \bibnamefont
  {von Nessi}}\ and\ \bibinfo {author} {\bibfnamefont {M.~J.}\ \bibnamefont
  {Hole}},\ }\href {\doibase http://dx.doi.org/10.1063/1.4811378} {\bibfield
  {journal} {\bibinfo  {journal} {Review of Scientific Instruments}\ }\textbf
  {\bibinfo {volume} {84}},\ \bibinfo {eid} {063505} (\bibinfo {year} {2013}),\
  http://dx.doi.org/10.1063/1.4811378}\BibitemShut {NoStop}%
\bibitem [{\citenamefont {Chilenski}\ \emph {et~al.}(2015)\citenamefont
  {Chilenski}, \citenamefont {Greenwald}, \citenamefont {Marzouk},
  \citenamefont {Howard}, \citenamefont {White}, \citenamefont {Rice},\ and\
  \citenamefont {Walk}}]{TSGP_CMOD}%
  \BibitemOpen
  \bibfield  {author} {\bibinfo {author} {\bibfnamefont {M.}~\bibnamefont
  {Chilenski}}, \bibinfo {author} {\bibfnamefont {M.}~\bibnamefont
  {Greenwald}}, \bibinfo {author} {\bibfnamefont {Y.}~\bibnamefont {Marzouk}},
  \bibinfo {author} {\bibfnamefont {N.}~\bibnamefont {Howard}}, \bibinfo
  {author} {\bibfnamefont {A.}~\bibnamefont {White}}, \bibinfo {author}
  {\bibfnamefont {J.}~\bibnamefont {Rice}}, \ and\ \bibinfo {author}
  {\bibfnamefont {J.}~\bibnamefont {Walk}},\ }\href
  {http://stacks.iop.org/0029-5515/55/i=2/a=023012} {\bibfield  {journal}
  {\bibinfo  {journal} {Nuclear Fusion}\ }\textbf {\bibinfo {volume} {55}},\
  \bibinfo {pages} {023012} (\bibinfo {year} {2015})}\BibitemShut {NoStop}%
\bibitem [{\citenamefont {Yamada}\ \emph {et~al.}(2003)\citenamefont {Yamada},
  \citenamefont {Narihara}, \citenamefont {Hayashi},\ and\ \citenamefont
  {Group}}]{RSI_LHDTS}%
  \BibitemOpen
  \bibfield  {author} {\bibinfo {author} {\bibfnamefont {I.}~\bibnamefont
  {Yamada}}, \bibinfo {author} {\bibfnamefont {K.}~\bibnamefont {Narihara}},
  \bibinfo {author} {\bibfnamefont {H.}~\bibnamefont {Hayashi}}, \ and\
  \bibinfo {author} {\bibfnamefont {L.~E.}\ \bibnamefont {Group}},\ }\href
  {\doibase http://dx.doi.org/10.1063/1.1538362} {\bibfield  {journal}
  {\bibinfo  {journal} {Review of Scientific Instruments}\ }\textbf {\bibinfo
  {volume} {74}},\ \bibinfo {pages} {1675} (\bibinfo {year}
  {2003})}\BibitemShut {NoStop}%
\bibitem [{\citenamefont {Yamada}\ \emph {et~al.}(2012)\citenamefont {Yamada},
  \citenamefont {Narihara}, \citenamefont {Funaba}, \citenamefont {Yasuhara},
  \citenamefont {Kohmoto}, \citenamefont {Hayashi}, \citenamefont {Hatae},
  \citenamefont {Tojo}, \citenamefont {Sakuma}, \citenamefont {Yoshida},
  \citenamefont {Fujita},\ and\ \citenamefont {Nakatsuka}}]{JINST_LHDTS}%
  \BibitemOpen
  \bibfield  {author} {\bibinfo {author} {\bibfnamefont {I.}~\bibnamefont
  {Yamada}}, \bibinfo {author} {\bibfnamefont {K.}~\bibnamefont {Narihara}},
  \bibinfo {author} {\bibfnamefont {H.}~\bibnamefont {Funaba}}, \bibinfo
  {author} {\bibfnamefont {R.}~\bibnamefont {Yasuhara}}, \bibinfo {author}
  {\bibfnamefont {T.}~\bibnamefont {Kohmoto}}, \bibinfo {author} {\bibfnamefont
  {H.}~\bibnamefont {Hayashi}}, \bibinfo {author} {\bibfnamefont
  {T.}~\bibnamefont {Hatae}}, \bibinfo {author} {\bibfnamefont
  {H.}~\bibnamefont {Tojo}}, \bibinfo {author} {\bibfnamefont {T.}~\bibnamefont
  {Sakuma}}, \bibinfo {author} {\bibfnamefont {H.}~\bibnamefont {Yoshida}},
  \bibinfo {author} {\bibfnamefont {H.}~\bibnamefont {Fujita}}, \ and\ \bibinfo
  {author} {\bibfnamefont {M.}~\bibnamefont {Nakatsuka}},\ }\href
  {http://stacks.iop.org/1748-0221/7/i=05/a=C05007} {\bibfield  {journal}
  {\bibinfo  {journal} {Journal of Instrumentation}\ }\textbf {\bibinfo
  {volume} {7}},\ \bibinfo {pages} {C05007} (\bibinfo {year}
  {2012})}\BibitemShut {NoStop}%
\bibitem [{\citenamefont {Bishop}(2006)}]{Bishop}%
  \BibitemOpen
  \bibfield  {author} {\bibinfo {author} {\bibfnamefont {C.}~\bibnamefont
  {Bishop}},\ }\href@noop {} {\emph {\bibinfo {title} {Pattern Recognition and
  Machine Learning}}}\ (\bibinfo  {publisher} {Springer-Verlag New York},\
  \bibinfo {year} {2006})\BibitemShut {NoStop}%
\bibitem [{\citenamefont {PEACOCK}\ \emph {et~al.}(1969)\citenamefont
  {PEACOCK}, \citenamefont {ROBINSON}, \citenamefont {FORREST}, \citenamefont
  {WILCOCK},\ and\ \citenamefont {SANNIKOV}}]{TS_T3}%
  \BibitemOpen
  \bibfield  {author} {\bibinfo {author} {\bibfnamefont {N.~J.}\ \bibnamefont
  {PEACOCK}}, \bibinfo {author} {\bibfnamefont {D.~C.}\ \bibnamefont
  {ROBINSON}}, \bibinfo {author} {\bibfnamefont {M.~J.}\ \bibnamefont
  {FORREST}}, \bibinfo {author} {\bibfnamefont {P.~D.}\ \bibnamefont
  {WILCOCK}}, \ and\ \bibinfo {author} {\bibfnamefont {V.~V.}\ \bibnamefont
  {SANNIKOV}},\ }\href {http://dx.doi.org/10.1038/224488a0} {\bibfield
  {journal} {\bibinfo  {journal} {Nature}\ }\textbf {\bibinfo {volume} {224}},\
  \bibinfo {pages} {488} (\bibinfo {year} {1969})}\BibitemShut {NoStop}%
\bibitem [{\citenamefont {Carlstrom}\ \emph {et~al.}(1992)\citenamefont
  {Carlstrom}, \citenamefont {Campbell}, \citenamefont {DeBoo}, \citenamefont
  {Evanko}, \citenamefont {Evans}, \citenamefont {Greenfield}, \citenamefont
  {Haskovec}, \citenamefont {Hsieh}, \citenamefont {McKee}, \citenamefont
  {Snider}, \citenamefont {Stockdale}, \citenamefont {Trost},\ and\
  \citenamefont {Thomas}}]{TS_DIIID}%
  \BibitemOpen
  \bibfield  {author} {\bibinfo {author} {\bibfnamefont {T.~N.}\ \bibnamefont
  {Carlstrom}}, \bibinfo {author} {\bibfnamefont {G.~L.}\ \bibnamefont
  {Campbell}}, \bibinfo {author} {\bibfnamefont {J.~C.}\ \bibnamefont {DeBoo}},
  \bibinfo {author} {\bibfnamefont {R.}~\bibnamefont {Evanko}}, \bibinfo
  {author} {\bibfnamefont {J.}~\bibnamefont {Evans}}, \bibinfo {author}
  {\bibfnamefont {C.~M.}\ \bibnamefont {Greenfield}}, \bibinfo {author}
  {\bibfnamefont {J.}~\bibnamefont {Haskovec}}, \bibinfo {author}
  {\bibfnamefont {C.~L.}\ \bibnamefont {Hsieh}}, \bibinfo {author}
  {\bibfnamefont {E.}~\bibnamefont {McKee}}, \bibinfo {author} {\bibfnamefont
  {R.~T.}\ \bibnamefont {Snider}}, \bibinfo {author} {\bibfnamefont
  {R.}~\bibnamefont {Stockdale}}, \bibinfo {author} {\bibfnamefont {P.~K.}\
  \bibnamefont {Trost}}, \ and\ \bibinfo {author} {\bibfnamefont {M.~P.}\
  \bibnamefont {Thomas}},\ }\href {\doibase
  http://dx.doi.org/10.1063/1.1143545} {\bibfield  {journal} {\bibinfo
  {journal} {Review of Scientific Instruments}\ }\textbf {\bibinfo {volume}
  {63}},\ \bibinfo {pages} {4901} (\bibinfo {year} {1992})}\BibitemShut
  {NoStop}%
\bibitem [{\citenamefont {Murmann}\ \emph {et~al.}(1992)\citenamefont
  {Murmann}, \citenamefont {G^^c3^^b6tsch}, \citenamefont {R^^c3^^b6hr},
  \citenamefont {Salzmann},\ and\ \citenamefont {Steuer}}]{TS_ASDEX}%
  \BibitemOpen
  \bibfield  {author} {\bibinfo {author} {\bibfnamefont {H.}~\bibnamefont
  {Murmann}}, \bibinfo {author} {\bibfnamefont {S.}~\bibnamefont
  {G^^c3^^b6tsch}}, \bibinfo {author} {\bibfnamefont {H.}~\bibnamefont
  {R^^c3^^b6hr}}, \bibinfo {author} {\bibfnamefont {H.}~\bibnamefont
  {Salzmann}}, \ and\ \bibinfo {author} {\bibfnamefont {K.~H.}\ \bibnamefont
  {Steuer}},\ }\href {\doibase http://dx.doi.org/10.1063/1.1143504} {\bibfield
  {journal} {\bibinfo  {journal} {Review of Scientific Instruments}\ }\textbf
  {\bibinfo {volume} {63}},\ \bibinfo {pages} {4941} (\bibinfo {year}
  {1992})}\BibitemShut {NoStop}%
\bibitem [{\citenamefont {LeBlanc}\ \emph {et~al.}(2003)\citenamefont
  {LeBlanc}, \citenamefont {Bell}, \citenamefont {Johnson}, \citenamefont
  {Hoffman}, \citenamefont {Long},\ and\ \citenamefont {Palladino}}]{TS_NSTX}%
  \BibitemOpen
  \bibfield  {author} {\bibinfo {author} {\bibfnamefont {B.~P.}\ \bibnamefont
  {LeBlanc}}, \bibinfo {author} {\bibfnamefont {R.~E.}\ \bibnamefont {Bell}},
  \bibinfo {author} {\bibfnamefont {D.~W.}\ \bibnamefont {Johnson}}, \bibinfo
  {author} {\bibfnamefont {D.~E.}\ \bibnamefont {Hoffman}}, \bibinfo {author}
  {\bibfnamefont {D.~C.}\ \bibnamefont {Long}}, \ and\ \bibinfo {author}
  {\bibfnamefont {R.~W.}\ \bibnamefont {Palladino}},\ }\href {\doibase
  http://dx.doi.org/10.1063/1.1532763} {\bibfield  {journal} {\bibinfo
  {journal} {Review of Scientific Instruments}\ }\textbf {\bibinfo {volume}
  {74}},\ \bibinfo {pages} {1659} (\bibinfo {year} {2003})}\BibitemShut
  {NoStop}%
\bibitem [{\citenamefont {Yamada}\ \emph {et~al.}(2007)\citenamefont {Yamada},
  \citenamefont {Narihara}, \citenamefont {Hayashi}, \citenamefont {Funaba},\
  and\ \citenamefont {experimental group}}]{PFR_LHDTS}%
  \BibitemOpen
  \bibfield  {author} {\bibinfo {author} {\bibfnamefont {I.}~\bibnamefont
  {Yamada}}, \bibinfo {author} {\bibfnamefont {L.}~\bibnamefont {Narihara}},
  \bibinfo {author} {\bibfnamefont {H.}~\bibnamefont {Hayashi}}, \bibinfo
  {author} {\bibfnamefont {H.}~\bibnamefont {Funaba}}, \ and\ \bibinfo {author}
  {\bibfnamefont {L.}~\bibnamefont {experimental group}},\ }\href
  {http://www.jspf.or.jp/PFR/PFR_articles/pfr2007S1/pfr2007_02-S1106.html}
  {\bibfield  {journal} {\bibinfo  {journal} {Plasma and Fusion Research}\
  }\textbf {\bibinfo {volume} {1}},\ \bibinfo {pages} {S1106} (\bibinfo {year}
  {2007})}\BibitemShut {NoStop}%
\bibitem [{\citenamefont {Suzuki}\ \emph {et~al.}(2013)\citenamefont {Suzuki},
  \citenamefont {Ida}, \citenamefont {Suzuki}, \citenamefont {Yoshida},
  \citenamefont {Emoto},\ and\ \citenamefont {Yokoyama}}]{LHD_MAP}%
  \BibitemOpen
  \bibfield  {author} {\bibinfo {author} {\bibfnamefont {C.}~\bibnamefont
  {Suzuki}}, \bibinfo {author} {\bibfnamefont {K.}~\bibnamefont {Ida}},
  \bibinfo {author} {\bibfnamefont {Y.}~\bibnamefont {Suzuki}}, \bibinfo
  {author} {\bibfnamefont {M.}~\bibnamefont {Yoshida}}, \bibinfo {author}
  {\bibfnamefont {M.}~\bibnamefont {Emoto}}, \ and\ \bibinfo {author}
  {\bibfnamefont {M.}~\bibnamefont {Yokoyama}},\ }\href
  {http://stacks.iop.org/0741-3335/55/i=1/a=014016} {\bibfield  {journal}
  {\bibinfo  {journal} {Plasma Physics and Controlled Fusion}\ }\textbf
  {\bibinfo {volume} {55}},\ \bibinfo {pages} {014016} (\bibinfo {year}
  {2013})}\BibitemShut {NoStop}%
\bibitem [{\citenamefont {Tresp}(2000)}]{BCM}%
  \BibitemOpen
  \bibfield  {author} {\bibinfo {author} {\bibfnamefont {V.}~\bibnamefont
  {Tresp}},\ }\href {\doibase doi: 10.1162/089976600300014908} {\bibfield
  {journal} {\bibinfo  {journal} {Neural Computation}\ }\textbf {\bibinfo
  {volume} {12}},\ \bibinfo {pages} {2719} (\bibinfo {year}
  {2000})}\BibitemShut {NoStop}%
\bibitem [{\citenamefont {Holland}\ \emph {et~al.}(2009)\citenamefont
  {Holland}, \citenamefont {White}, \citenamefont {McKee}, \citenamefont
  {Shafer}, \citenamefont {Candy}, \citenamefont {Waltz}, \citenamefont
  {Schmitz},\ and\ \citenamefont {Tynan}}]{SIMULATION1}%
  \BibitemOpen
  \bibfield  {author} {\bibinfo {author} {\bibfnamefont {C.}~\bibnamefont
  {Holland}}, \bibinfo {author} {\bibfnamefont {A.~E.}\ \bibnamefont {White}},
  \bibinfo {author} {\bibfnamefont {G.~R.}\ \bibnamefont {McKee}}, \bibinfo
  {author} {\bibfnamefont {M.~W.}\ \bibnamefont {Shafer}}, \bibinfo {author}
  {\bibfnamefont {J.}~\bibnamefont {Candy}}, \bibinfo {author} {\bibfnamefont
  {R.~E.}\ \bibnamefont {Waltz}}, \bibinfo {author} {\bibfnamefont
  {L.}~\bibnamefont {Schmitz}}, \ and\ \bibinfo {author} {\bibfnamefont
  {G.~R.}\ \bibnamefont {Tynan}},\ }\href {\doibase
  http://dx.doi.org/10.1063/1.3085792} {\bibfield  {journal} {\bibinfo
  {journal} {Physics of Plasmas}\ }\textbf {\bibinfo {volume} {16}},\ \bibinfo
  {eid} {052301} (\bibinfo {year} {2009}),\
  http://dx.doi.org/10.1063/1.3085792}\BibitemShut {NoStop}%
\bibitem [{\citenamefont {Edwin V.~Bonilla}(2007)}]{MultitaskGP}%
  \BibitemOpen
  \bibfield  {author} {\bibinfo {author} {\bibfnamefont {C.~K. I.~W.}\
  \bibnamefont {Edwin V.~Bonilla}, \bibfnamefont {Kian Ming A.~Chai}},\ }\href
  {http://citeweb.info/20080008055} {\bibfield  {journal} {\bibinfo  {journal}
  {Proceedings of NIPS2007}\ } (\bibinfo {year} {2007})}\BibitemShut {NoStop}%
\end{thebibliography}%

\end{document}